\documentclass{jetpl}

\twocolumn

\usepackage{amsmath,amssymb,epsfig,color,pstricks,bm,graphics,alltt}

\begin{document}

\lat

\title{Multiple Bands -- A Key to High - Temperature Superconductivity
in Iron Arsenides?}

\rtitle{Multiple Bands -- A Key to High - Temperature Superconductivity?}

\sodtitle{Multiple Bands -- A Key to High - Temperature Superconductivity
in Iron Arsenides?}

\author{E.\ Z.\ Kuchinskii \thanks{E-mail: kuchinsk@iep.uran.ru}, M.\ V.\ Sadovskii \thanks{E-mail: sadovski@iep.uran.ru}}

\rauthor{E.\ Z.\ Kuchinskii, M.\ V.\ Sadovskii}

\sodauthor{Kuchinskii, Sadovskii }

\sodauthor{Kuchinskii, Sadovskii }

\address{Institute for Electrophysics, Russian Academy of Sciences, 
Ural Division, 620016 Ekaterinburg, Russia}


\abstract{
In the framework of four-band model of superconductivity in iron
arsenides proposed by Barzykin and Gor'kov we analyze the gap ratios on
hole - like and electron - like Fermi - surface cylinders. It is shown that
experimentally observed (ARPES) gap ratios can be obtained only within rather
strict limits on the values of pairing coupling constants.
The difference of $T_c$ values in 1111 and 122 systems is reasonably explained
by the relative values of partial densities of states. The multiple bands
electronic structure of these systems leads to a significant enhancement of
effective pairing coupling constant determining $T_c$, so that high
enough $T_c$ values can be achieved even for the case of rather small
intraband and interband pairing interactions.
}


\maketitle

The discovery of high -- temperature superconductivity in layered FeAs
compounds stimulated active experimental and theoretical studies of these
new superconductors \cite{UFN_90}. The main anomaly of these systems is 
their multiple bands nature. Electronic structure in a narrow enough energy
interval around the Fermi level is formed almost only from the $d$ - states 
of Fe. In fact, electronic spectrum of iron arsenides was calculated in a 
number of papers \cite{dolg,mazin,Xu1282,Shein,Singh2643}. 
The Fermi surface consists of several hole - like and electron - like cylinders 
and on each of these its ``own'' superconducting gap can be formed. In the 
energy interval relevant to superconductivity electronic spectrum is especially
simple \cite{Nek1239,Nek2630,Nek1010}. It was used by Barzykin and Gor'kov to
formulate a simple (analytic) model of superconducting state of new
superconductors \cite{Gork}. 

\begin{figure}[!h]
\includegraphics[clip=true,width=0.7\columnwidth]{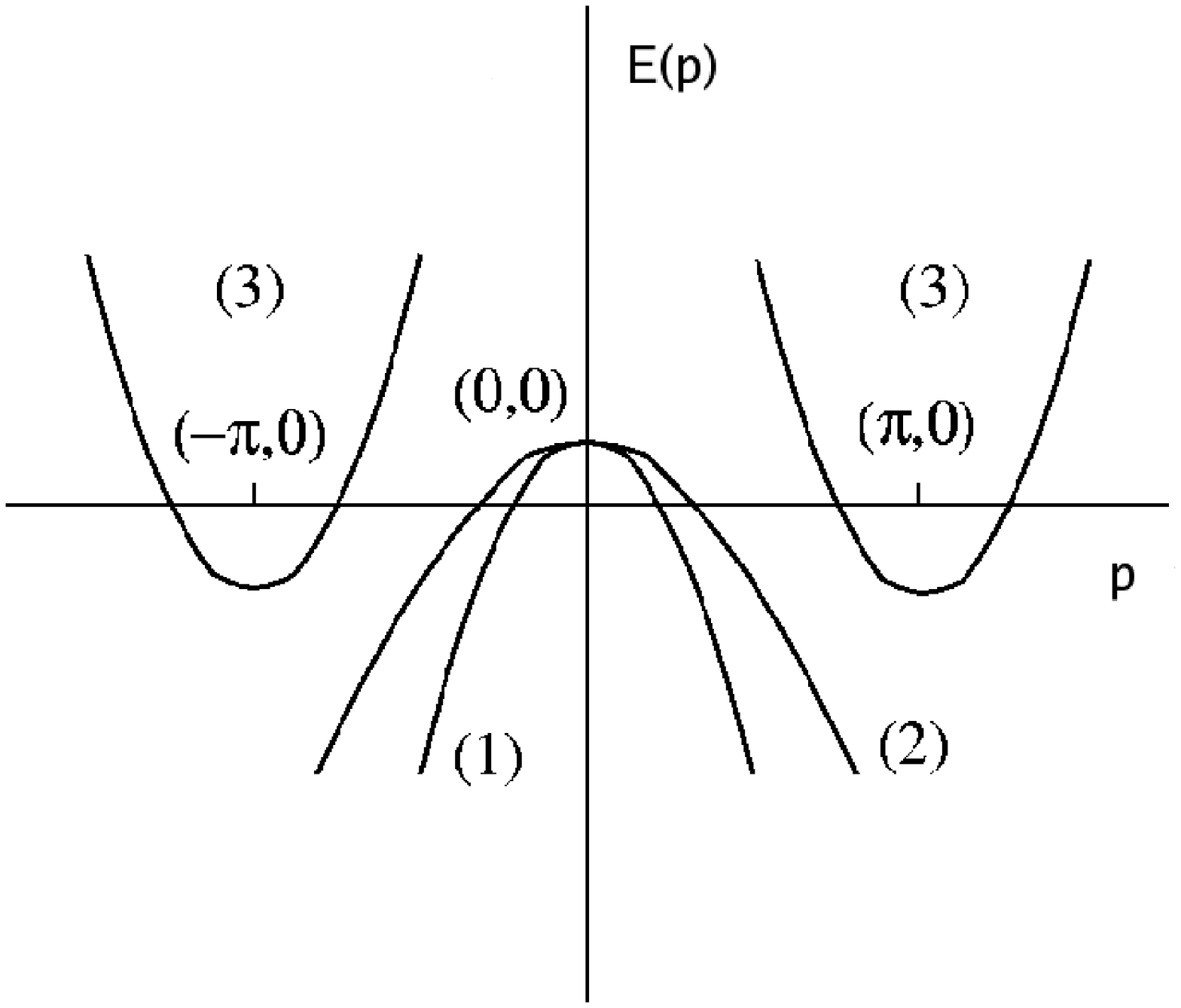}
\includegraphics[clip=true,width=0.7\columnwidth]{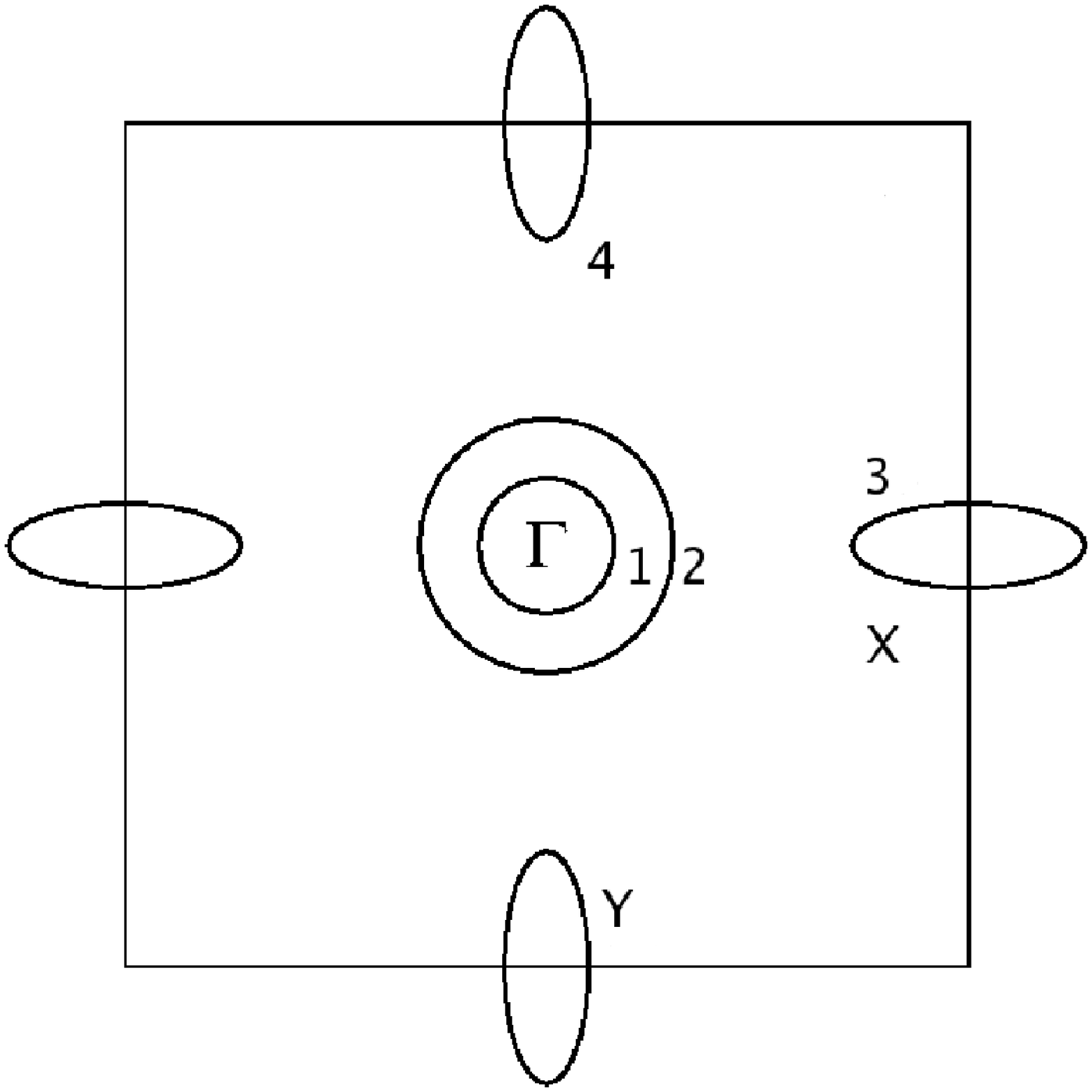}
\caption{Fig. 1. Schematic electronic spectrum and Fermi surfaces 
of FeAs superconductor in the extended band picture. There are two hole - like
cylinders around point $\Gamma$, while electron - like cylinders are around
$X$ ($Y$) points \cite{Gork}.} 
\label{barz_gork} 
\end{figure}

Schematically, the simplified electronic spectrum and Fermi surfaces of these 
systems are shown in Fig. \ref{barz_gork} \cite{Gork}. There are two hole - like
Fermi surface cylinders surrounding the $\Gamma$ point and two electronic
pockets around  $X$ and $Y$  points in extended Brillouin zone.

Let $\Delta_i$ be a superconducting order - parameter (gap) on the 
$i$-th sheet of the Fermi surface. The value of $\Delta_i$ is
determined by self -- consistency equation for the anomalous Gor'kov
Green's function.

Pairing BCS -- like interaction can be represented by a matrix:
\begin{equation}
V=\left(\begin{array}{cccc} 
u & w & t & t \\
w & u' & t & t \\ 
t & t & \lambda & \mu \\
t & t & \mu & \lambda 
\end{array}\right).
\label{mmatr}
\end{equation}
where matrix elements $V^{i,j}$ define intraband and interband pairing coupling
constants. For example, $\lambda = V^{eX,eX} = V^{eY,eY}$ determines pairing
interactions on the same electronic pocket at point $X$ or $Y$, $\mu = V^{eX,eY}$ 
connects electrons of different pockets at these points, 
$u = V^{h1,h1}$, $u'=V^{h2,h2}$ and $w=V^{h1,h2}$ characterize BCS interactions
within two hole -- like pockets --- the small one ($h1$) and the large one
($h2$), as well as between these pockets, while $t = V^{h,eX}=V^{h,eY}$ 
couple electrons at points $X$ and $\Gamma$. In Ref. \cite{Gork} it was
assumed that  $u=u'=w$. This assumption seems to be too strong and below we
analyze the general case.

Superconducting critical temperature  $T_c$ is determined by an effective 
pairing coupling constant $g_{eff}$:
\begin{equation}
T_c = \frac{2 \gamma \omega_c}{\pi}\,e^{- 1/g_{eff}}, 
\label{TC1} 
\end{equation}
where $\omega_c$ is the usual cut -- off frequency in Cooper channel (assumed to
be the same for all types of couplings under consideration -- a simplification!), 
while $g_{eff}$ in this model is defined by the solution of the system of
linearized gap equations:
\begin{equation} 
g_{eff}\Delta_{i}  =   \sum_{j}g_{ij}\Delta_{j} \,,
\label{gapeqlin} 
\end{equation}
where 
\begin{equation}
g_{ij} \equiv  - V^{i,j} \nu_{j},\quad g_{eff}^{-1}=\ln \frac{2\gamma}{\pi}\frac{\omega_c}{T_c}.
\label{intmatr}
\end{equation}
The matrix of dimensionless coupling constants $g_{ij}$ is determined by 
matrix elements of (\ref{mmatr}) and partial densities of states on
different Fermi surface cylinders --- $\nu_j$ is density of states per single 
spin projection on the $j$-th cylinder.

From symmetry it is clear that $\nu_3=\nu_4$ and the system (\ref{gapeqlin}) 
possesses solutions of two types \cite{Gork}:

1) solution corresponding to $d_{x^2 - y^2}$ symmetry, when gaps on different
pockets at points $X$ and $Y$ differ by sign, while gaps on hole - like
pockets are just zero:

\begin{equation}
\Delta_1 = \Delta_2 = 0, \ \ 
\Delta_3 = - \Delta_4 = \Delta,
\end{equation}

and

\begin{equation}
g_{eff} = (\mu - \lambda) \nu_3. 
\label{dwve}
\end{equation}

2) solutions corresponding to the so called $s^{\pm}$ pairing \cite{mazin}, 
for which gaps on the cylinders at points $X$ and $Y$ are equal to each other:
$\Delta_3 =  \Delta_4$, while gaps on Fermi surfaces surrounding the point
$\Gamma$ are of different sign in case of repulsive interaction between
electron - like and hole - like pockets ($t>0$), and of the same sign for the
case of $t<0$.

As in this case we have $\Delta_3 =  \Delta_4$ and $\nu_3=\nu_4$, 
two equations in (\ref{gapeqlin}) just coincide and instead of 
(\ref{mmatr}), (\ref{intmatr}) we are dealing with $3\times 3$ matrix 
of coupling constants of the following form:
\begin{equation}
-\hat g=\left(\begin{array}{ccc} 
u\nu_1 & w\nu_2 & 2t\nu_3 \\
w\nu_1 & u'\nu_2 & 2t\nu_3 \\ 
t\nu_1 & t\nu_2 & 2\bar \lambda \nu_3  
\end{array}\right),
\label{effmatr}
\end{equation}
where $\bar \lambda =\frac{\lambda +\mu }{2}$ and (\ref{gapeqlin})
reduces to the standard problem of finding eigenvalues and eigenvectors 
for the matrix of dimensionless couplings $g_{ij}$ (\ref{effmatr}), which
has three solutions, determined by cubic secular equation:
\begin{equation}
Det(g_{ij}-g_{eff}\delta _{ij})=0
\label{solvgeff}
\end{equation}
Physical solution corresponds to a maximal positive value of $g_{eff}$, 
which determines the highest value of $T_c$.

Under the simple assumption of Ref. \cite{Gork}, when $u=u'=w$, situation
simplifies further, as in (\ref{gapeqlin}) only two independent equations
remain, so that we have $2\times 2$  matrix of coupling constants and
(\ref{solvgeff}) reduces to a quadratic equation. Then we easily obtain
\cite{Gork}:
\begin{equation}
\Delta_1 = \Delta_2 = \kappa \Delta , \ \ \Delta_3 = \Delta_4 = \Delta,
\label{swaveord}
\end{equation}
where $\kappa^{-1} = - (g_{eff} + u (\nu_1 + \nu_2))/(t \nu_3)$, and
maximal effective pairing constant is given by:
\begin{eqnarray}
\label{swaveg}
2 g_{eff} &=& - u (\nu_1 + \nu_2) - 2\bar \lambda \nu_3 + \\
& & + \sqrt{(u (\nu_1 + \nu_2) - 2\bar \lambda \nu_3)^2 + 8 t^2 \nu_3 
(\nu_1 + \nu_2)} \nonumber  
\end{eqnarray}
Possibility of $s^{\pm}$ -- pairing in FeAs compounds was first noted in
Ref. \cite{mazin}. This kind of solution qualitatively agrees with ARPES data
of Refs. \cite{Ding0419,Wray2185,Evt4455}, except the result
$\Delta_1 = \Delta_2$ (\ref{swaveord}), which contradicts the established
experimental fact --- the gap on the small hole - like cylinder $\Delta_1$ is
approximately twice as large as the gap $\Delta_2$ on the large cylinder.
In fact, this contradiction is basically due to an unnecessary limitation to
the case of $u = u'=w$ used in Ref. \cite{Gork}.

The system of linearized gap equations determines their ratios on different
sheets of the Fermi surface for temperatures $T\to T_c$. In general case, 
the temperature dependence of gaps is determined by the generalized BCS
equations:
\begin{equation} 
\Delta_i =  \sum_{j} g_{ij}\Delta_j \int_{0}^{\omega_c}d\xi 
\frac{th\frac{\sqrt{\xi^2+\Delta_j^2}}{2T}}{\sqrt{\xi^2+\Delta_j^2}} ,
\label{gapeq_T} 
\end{equation}
For $T\to 0$ these equations take the form:
\begin{equation}
\Delta_i =  \sum_{j} g_{ij}\Delta_jF\left( \frac{\Delta_j}{\omega_c}\right),
\label{gapeq_T0}
\end{equation}
where we have introduced $F(x)=ln\left( \frac{1+\sqrt{1+x^2}}{|x|}\right)$.

Below we present the results of numerical studies of Eqs. (\ref{gapeqlin}) 
and (\ref{gapeq_T0}) for typical values of parameters (couplings).

Let us denote the pairing coupling constant on a small hole - like cylinder as
$g=g_{11}$. In the following we take $g=0.2$, which allows us to remain within 
the limits of weak coupling approximation.

The ratio of partial densities of states for different Fermi surface cylinders
in quasi - two - dimensional case can be approximated by effective mass ratio
on the same cylinders. These can be estimated from the data for electronic
dispersions in symmetric directions in the Brillouin zone, obtained in LDA
calculations \cite{Nek1239,Nek2630,Nek1010}. For REOFeAs series 
(RE=La,Ce,Nd,Pr,Sm...) (1111) and for BaFe$_2$As$_2$ (122) from these data we
get:
\begin{equation}
\begin{array}{cl}
\frac{\nu_2}{\nu_1}\approx 1.18, \qquad \frac{\nu_3}{\nu_1}\approx 0.64 & \qquad \mbox{for 1111} \\ 
\frac{\nu_2}{\nu_1}\approx 1.26, \qquad \frac{\nu_3}{\nu_1}\approx 0.34 & \qquad \mbox{for 122.}
\end{array}
\label{dosratio}
\end{equation}

We suppose that pairing interactions on hole - like cylinders and between them,
as well as on electron - like cylinders and between them, are most probably
determined by electron - phonon interaction, the relevance of which is
clearly demonstrated by rather strong isotope effect, observed in Ref. 
\cite{Liu2694}. At the same time, interband pairing interaction between hole -
like and electron - like cylinders is probably due to antiferromagnetic
fluctuations and is repulsive ($t>0$). It should be noted that parameter
$t$ from coupling constants matrix (\ref{effmatr}) enters Eq. (\ref{solvgeff}), 
determining $g_{eff}$, only via $t^2$, i.e. independent of sign. Thus its sign
does not change the value of an effective pairing coupling constant
and that of $T_c$. Repulsion between quasiparticles on hole - like and 
electron - like cylinders does not suppress, but actually enhances
superconductivity leading to the increase of $g_{eff}$. Also the sign change
of $t$ does not change the absolute values of gaps on different cylinders,
though the repulsion between electron - like and hole - like cylinders
($t>0$) leads to different signs of gaps at these cylinders, while for 
the case of $t<0$ both gaps acquire the same sign.

\begin{figure}[!h]
\includegraphics[clip=true,width=\columnwidth]{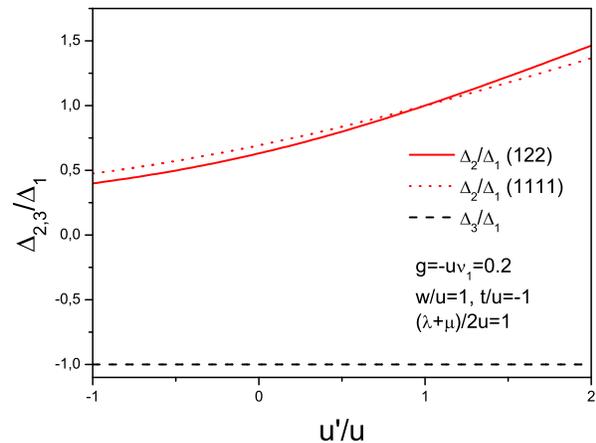}
\caption{Fig. 2. Dependence of gap ratios on different pockets of the Fermi
surface on $u'/u$ for $g=0.2$, $w/u=1$, $t/u=-1$, $\bar \lambda /u=1$ 
and partial density of states ratios given by (\ref{dosratio}). } 
\label{gapratio} 
\end{figure}

Despite rather large number of free parameters of the model it is not easy to
obtain the observable (in ARPES experiments of Refs. 
\cite{Ding0419,Wray2185,Evt4455}) values of the ratios 
$|\Delta_2/\Delta_1|\approx 0.5$ and $|\Delta_3/\Delta_1|\approx 1$. In fact
it requires small enough attraction (or even repulsion, $u'>0$) on the 
``large'' hole - like cylinder (cf. Fig.\ref{gapratio}). In the following 
we assume the ratios of pairing coupling constants as $w/u=1$, $t/u=-1$, 
$\bar \lambda /u=1$, which guarantees us the ratio $|\Delta_3/\Delta_1|=1$ 
for any values of $u'$  and arbitrary ratios of partial densities of states
at different cylinders. Another choice of pairing couplings producing 
$|\Delta_3/\Delta_1|=1$ is also possible, but in general we need larger
repulsion on ``large'' hole - like cylinder to get
$|\Delta_2/\Delta_1|\approx 0.5$. In Fig. \ref{gapratio} we show the
dependences of the gap ratios at $T=0$ on $u'/u$, obtained from (\ref{gapeq_T0}), 
using the partial density of states ratios on different cylinders (\ref{dosratio}), 
characteristic for (1111) and (122) systems. The gap ratios for $T\to T_c$ 
differ from the values obtained at $T=0$ rather insignificantly.

In Ref. \cite{dolgmazin} a two - band model with two hole - like cylinders
was analyzed, assuming that only interband coupling exists, i.e. the
coupling constants matrix has the form:
\begin{equation}
-g_{ij}=\left(\begin{array}{cc} 
0 & w\nu_2 \\
w\nu_1 & 0   
\end{array}\right).
\label{effmatr_dolgov}
\end{equation}
Under this assumption the gap ratio on hole - like cylinders is given by:
\begin{equation}
\frac{\Delta_2}{\Delta_1}=\sqrt{\frac{\nu_1}{\nu_2}}
\label{gaps_dolgov}
\end{equation}
so that for characteristic for $BaFe_2As_2$ value of $\nu_2/\nu_1\approx 1.26$ 
we obtain $\Delta_1/\Delta_2\approx 1.12$, which is significantly lower than the
experimentally observed value of \cite{Ding0419} $\Delta_1/\Delta_2\approx 2$.

Four - band model somehow similar to that considered above was analyzed in
Ref. \cite{DiCastro}, where temperature dependences of gaps (with proper ratios)
on different sheets of the Fermi surface were calculated along with the 
temperature dependence of superfluid electron density. However, in this work 
no analysis was made of the important role of multiple bands structure for the 
increase of $T_c$, which we shall discuss shortly.

\begin{figure}[!h]
\includegraphics[clip=true,width=\columnwidth]{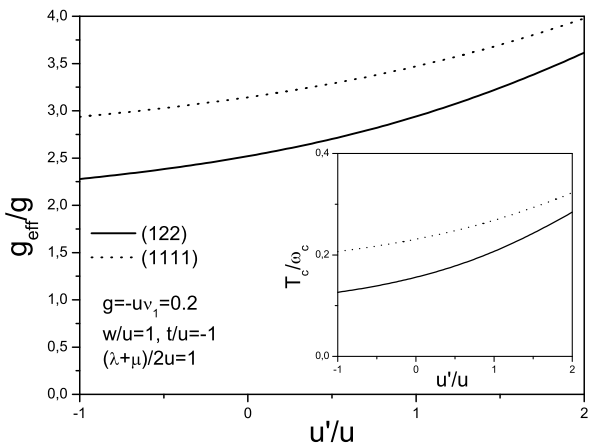}
\caption{Fig. 3. Dependence of effective pairing coupling constant on
$u'/u$ for $g=0.2$, $w/u=1$, $t/u=-1$, $\bar \lambda /u=1$ and partial density 
of states ratios on different Fermi surface pockets given by (\ref{dosratio}). 
At the insert --- similar dependence of the critical temperature.} 
\label{geff} 
\end{figure}

In Fig.\ref{geff} we show the dependence of an effective pairing coupling
constant and superconducting critical temperature on $u'/u$ for both classes
of FeAs systems (1111 and 122). It is clearly seen that the effective coupling
constant $g_{eff}$ is significantly larger than the pairing constant $g$ on
the small hole - like cylinder. It can be said that coupling constants from
different cylinders effectively produce ``additive'' effect. 
In fact this can lead to high enough
values of $T_c$ even for relatively small values of intraband \cite{dolg} and
interband pairing constants. Actually, using this type of estimates we can
convince ourselves that the critical temperature for superconducting transition
with $d_{x^2 - y^2}$ gap symmetry, which is determined by an effective pairing
constant given by (\ref{dwve}), is always smaller (for typical values of 
parameters) than the critical temperature for $s^{\pm}$ pairing.

To clarify the reasons for the growth of effective pairing coupling it is
helpful to analyze the most simple case, when all pairing interactions 
(both intraband and interband) in (\ref{mmatr}) are just the same (and equal e.g.
to $u$), and all partial densities of states on all four Fermi surface pockets
are also the same (and equal e.g. to $\nu_1$). In this case we obtain
$g_{eff}=4g=-4u\nu_1$, which simply corresponds to the fact that now the total 
density of states at the Fermi level is four times partial. However, in real
situation the growth of an effective pairing constant does not reduce to this
simple summation of partial densities of states. In particular, the effective
pairing coupling may be much larger than the simple sum of intraband (diagonal)
dimensionless coupling constants, e.g. in case of significant interband
pairing interaction, which can be present in iron arsenides, where the pairing
interaction between electron - like and hole - like cylinders is most probably
attributed to antiferromagnetic fluctuations.

It can be estimated that with the same values of interaction constants in
(\ref{mmatr}) the critical temperature in 1111 - type systems is typically
larger than in 122 just due to the difference of partial densities of states
as given in (\ref{dosratio}) (cf. insert in Fig.\ref{geff}). 
For example, in case of $u'/u=0$ 
(with the values of parameters for 122 - system we get the ratio of gaps 
$\Delta_2/\Delta_1\approx 0.6$) the calculated ratio of critical temperatures of 
122 and 1111 systems $T_c(122)/T_c(1111)=0.67$ is very close to the observed
ration of maximal critical temperatures obtained for these systems:
$38 K/55 K\approx 0.69$. Thus the typical difference of $T_c$'s for both
classes of new superconductors can be attributed to the different values of 
partial densities of states on corresponding Fermi surface cylinders, 
despite the fact that total densities of states at the Fermi level in these
systems are practically the same \cite{Nek1239,Nek2630,Nek1010}. Of course, 
the accuracy obtained should not be taken too seriously, as in real systems 
rather strong renormalization effects of electronic spectrum (effective masses,
bandwidths etc.) in comparison with the results of LDA calculations are 
definitely present (and observed in ARPES experiments), e.g. due to moderate or 
probably even strong enough Coulomb correlations \cite{UFN_90}. The main
conclusion following from our analysis is the simple fact that the value of
$T_c$ in multiple bands systems is determined by the relations between partial
densities of states on different sheets of the Fermi surface, not by the total
density of states at the Fermi level as in the standard BCS model. 

It should be noted that for the first time (though only implicitly)
the role of multiple bands structure of electronic spectrum as the reason for
the increase of superconducting $T_c$ was apparently discussed in relation
to superconductivity in multivalley doped semiconductors \cite{Fir,Coh}.
In these works the important role of interband electron - phonon pairing
mechanism was also stressed. It was noted that such processes with large 
momentum transfer, leading to reduced screening, may be most relevant for the
increase of $T_c$. This fact can be also important for new superconductors 
besides the abovementioned role of pairing due to spin fluctuations.

Direct experimental confirmation of the role of multiple bands in new
superconductors follows from ARPES measurements on extremely (hole) overdoped
system KFe$_2$As$_2$ with $T_c$=3K \cite{Sato} and similar heavily (electron)
overdoped BaFe$_{1.7}$Co$_{0.3}$As$_2$ \cite{Sekiba}, where superconductivity
is just absent. From these measurements it is clearly seen how the
disappearance of electronic pockets in the first system and hole - like pockets
in the second one leads to strong suppression or even the complete
disappearance of $T_c$.

\begin{figure}[!h]
\includegraphics[clip=true,width=\columnwidth]{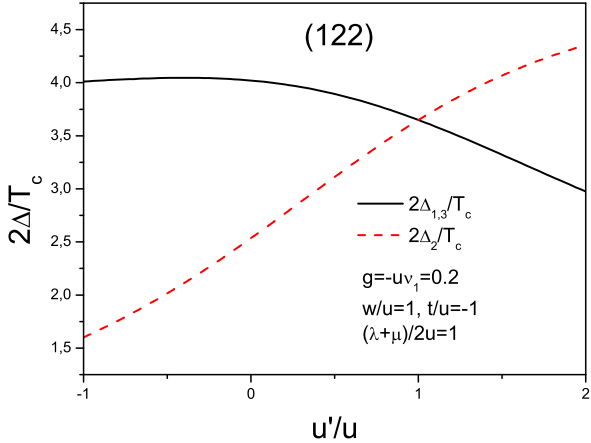}
\caption{Fig. 4. Dependence of $\frac{2\Delta}{T_c}$ ratio on $u'/u$ 
for 122 - system with $g=0.2$, $w/u=1$, $t/u=-1$, $\bar \lambda /u=1$ 
and partial densities of states ratios on different Fermi surface
sheets, as given in (\ref{dosratio}).} 
\label{2gaptc} 
\end{figure}

To conclude, on Fig. \ref{2gaptc} we show the dependence of
$\frac{2\Delta}{T_c}$ ratio on different sheets of the Fermi surface on
the ratio of coupling constants $u'/u$. Here it is important to note that the
value of this characteristic ratio can be significantly different from the
standard BCS value $\frac{2\Delta}{T_c}\approx 3.5$. However, the values shown
in Fig. \ref{2gaptc} are much lower than the ratios observed in ARPES experiments
\cite{Ding0419,Wray2185,Evt4455}, where the typical values are
$\frac{2\Delta_{1,3}}{T_c}\approx 7.5$ and $\frac{2\Delta_2}{T_c}\approx 3.7$,
which is apparently due to the strong coupling effects important in real systems.
Our analysis was limited to the standard BCS - like weak coupling approach.
Strong coupling Eliashberg -- type analysis of multiple bands effects for 
new superconductors is yet to be done. Preliminary results on gap ratios
in the strong coupling limit for the simple two - band model were derived 
in Ref. \cite{dolgmazin}.  

Authors are grateful to L.P. Gor'kov and E.G. Maksimov for stimulating
discussions.

This work is partly supported by RFBR grant 08-02-00021 and was performed
within the framework of programs of fundamental research of the Russian Academy
of Sciences (RAS) ``Quantum physics of condensed matter'' and of the Physics 
Division of RAS  ``Strongly correlated electrons in solid states''. 



\end{document}